\documentclass[twocolumn,showpacs,prl]{revtex4}

\usepackage{amsmath,amsfonts,amssymb,bbm}
\usepackage{txfonts}
\usepackage{type1cm}

\newcommand{\id}{\mathbbm{1}}
\newcommand{\eq}[1]{Eq.~(\ref{#1})}

\begin{document}

\bibliographystyle{apsrev}

\title{Monogamy inequality for distributed Gaussian entanglement}

\author{Tohya Hiroshima$^{1}$, Gerardo Adesso$^{2,3}$, and Fabrizio Illuminati$^{2}$}
\affiliation{
$^{1}$ Quantum Computation and Information Project, ERATO-SORST, Japan Science and Technology Agency,\\
Daini Hongo White Building 201, Hongo 5-28-3, Bunkyo-ku, Tokyo 113-0033, Japan. \\
$^{2}$ Dipartimento di Fisica ``E. R. Caianiello,'' Universit\`{a} degli Studi di Salerno,
CNR-Coherentia, Gruppo di Salerno, \\
and INFN Sezione di Napoli-Gruppo Collegato di Salerno, Via S. Allende, 84081 Baronissi, SA, Italy. \\
$^{3}$ Centre for Quantum Computation, DAMTP, Centre for Mathematical Sciences, University of Cambridge,\\
Wilberforce Road, Cambridge CB3 0WA, United Kingdom.}

\date{November 4, 2006}
%%%%%%%%%%%%%%%%%%%%%%%%%%%%%%%%%%%%%%%%%%%%%%%%%%%%%%%%%%%%%%%%%%%%%%%%%%%%%%%
\begin{abstract}
We show that for all $n$-mode Gaussian states of continuous variable
systems, the  entanglement shared among $n$ parties exhibits the
fundamental monogamy property. The monogamy inequality is proven by
introducing the Gaussian tangle, an entanglement monotone under
Gaussian local operations and classical communication, which is
defined in terms of the squared negativity in complete analogy with
the case of $n$-qubit systems. Our results elucidate the structure
of quantum correlations in many-body harmonic lattice systems.
\end{abstract}
%%%%%%%%%%%%%%%%%%%%%%%%%%%%%%%%%%%%%%%%%%%%%%%%%%%%%%%%%%%%%%%%%%%%%%%%%%%%%%%
\maketitle
%%%%%%%%%%%%%%%%%%%%%%%%%%%%%%%%%%%%%%%%%%%%%%%%%%%%%%%%%%%%%%%%%%%%%%%%%%%%%%%

Quantum entanglement, at the heart of quantum correlations and a
direct consequence of the superposition principle, cannot be freely
shared among many parties, unlike classical correlations. This is
the so-called {\it monogamy} property \cite{Ter} and is one of the
fundamental traits of entanglement and of quantum mechanics itself
\cite{Monogamy}. Seminal observations on the monogamy property and
its precise quantitative statement in mathematical terms are due to
Coffman, Kundu, and Wootters \cite{CKW}. They proved the following
inequality for a generic state $\rho _{ABC}$ of three qubits,
\begin{equation} \label{eq:CKW}
\tau (\rho _{A:BC})\, \geq \, \tau (\rho _{A:B})\, + \, \tau (\rho _{A:C}) \; ,
\end{equation}
where $\tau $ is an entanglement monotone \cite{Vid} known as the
{\it tangle} \cite{CKW}, $\tau (\rho _{A:BC})$ stands for the
bipartite entanglement across the bipartition $A:BC$, and $\rho
_{A:B(C)}=\mathrm{Tr} _{C(B)} \{ \rho _{ABC} \}$.
Inequality~(\ref{eq:CKW}) clearly elucidates the restriction on the
sharing of entanglement among the three parties.
The monogamy inequality~(\ref{eq:CKW}) holds for the
tangle defined through the square of the concurrence \cite{Hil, Woo}.
%%Despite its fundamental nature, the monogamy
%%inequality~(\ref{eq:CKW}) is satisfied by the tangle defined through
%%the square of concurrence \cite{Hil,Woo} but can be violated if
%%$\tau$ is replaced by other entanglement measures such as the
%%concurrence itself or the entanglement of formation \cite{Woo,BDSW}.
%%This indicates that the proper choice of the bipartite entanglement
%%monotone is of primal importance to capture the monogamous nature of
%%entanglement and to address the {\it a priori} trade-off on
%%entanglement distribution in the elegant form of
%%Inequality~(\ref{eq:CKW}).
%%but fails if $\tau$ is replaced by other entanglement measures such as
%%the concurrence itself or the entanglement of formation
%%\cite{Woo,BDSW}.
%%This indicates that the proper choice of the
%%bipartite entanglement monotone is of crucial importance to capture
%%the monogamous nature of entanglement and to address the trade-off
%%on entanglement distribution in the elegant form of
%%Inequality~(\ref{eq:CKW}) \cite{AImono}.
Recently, Osborne and Verstraete \cite{OV} have generalized
the monogamy inequality to $n$-qubit systems, proving a longstanding
conjecture formulated in Ref.~\cite{CKW}, with important consequences
for the description of the entanglement structure in many-body spin
systems.

For higher-dimensional systems much less is known on the
qualification, let alone the quantification of bipartite and
multipartite entanglement, the situation worsening with increasing
dimension of the Hilbert space due to the exponential increase in
the complexity of states. Remarkably, in the limit of
continuous variable (CV) systems with an infinite dimensional
Hilbert space, if one focuses on the theoretically and practically
relevant class of Gaussian states an almost as comprehensive
characterization of entanglement has been achieved as in the case of
qubit systems \cite{BL}. In this context,
%%a crucial piece of
%%knowledge for the understanding of CV entanglement and its
%%distribution is still missing. Namely,
the natural question arises
whether the monogamy inequality holds as well for entanglement
sharing in CV systems, and in particular in generic Gaussian states
of harmonic lattices. The first step towards answering this question
has been taken by Adesso and Illuminati \cite{AI} (see also
\cite{ASI06}). They proved the monogamy inequality for arbitrary
three-mode Gaussian states and for symmetric $n$-mode Gaussian
states, defining the CV tangle or ``contangle'' as the square of the
logarithmic negativity (an entanglement monotone \cite{VW}).
However, it is known \cite{LCOK} that in two-qubit systems the
concurrence is equivalent to another  related entanglement measure,
the negativity \cite{ZPSL}, so it appears natural to promote the
tangle to CV systems by defining it in terms of the squared
negativity itself \cite{notelogneg}.

In this paper we provide the complete answer to the question posed
above. We prove that the monogamy inequality {\it does} hold for all
Gaussian states of multimode CV systems with an arbitrary number $n$
of modes and parties $A_{1},\ldots,A_{n}$, thus generalizing the
results of \cite{AI,ASI06}. As a measure of bipartite entanglement,
we define the Gaussian tangle via the square of negativity, in
direct analogy with the case of $n$-qubit systems \cite{OV}. Our
proof is based on the symplectic analysis of covariance matrices and
on the properties of Gaussian measures of entanglement \cite{AI05}.
The monogamy constraint has important implications on the structural
characterization of entanglement sharing in CV systems
\cite{AI,ASI06}, in the context of entanglement frustration in
harmonic lattices \cite{frusta}, and for practical applications such
as secure key distribution and communication networks with
continuous variables.

For a $A_{1}:A_{2}\ldots A_{n}$ bipartition associated to a pure
Gaussian state $\rho _{A:B}^{(p)}$ with $A=A_{1}$ (a subsystem of a
single mode) and $B=A_{2}\ldots A_{n}$, we define the following
quantity
\begin{equation} \label{eq:G-tangle_pure}
\tau _{G}(\rho _{A:B}^{(p)})=\mathcal{N}^{2}(\rho _{A:B}^{(p)}).
\end{equation}
Here, $\mathcal{N}(\rho)=\left( \left\| \rho ^{T_{A}}\right\|
_{1}-1\right) /2$ is the negativity \cite{VW,ZPSL}, $\left\| \cdot
\right\| _{1}$ denotes the trace norm, and $\rho ^{T_{A}}$ stands
for the partial transposition of $\rho $ with respect to the
subsystem $A$. The functional $\tau _{G}$, like the negativity
$\mathcal{N}$, vanishes on separable states and does not increase
under local operations and classical communication (LOCC), i.e., it
is a proper measure of pure-state bipartite entanglement \cite{Vid}.
It can be naturally extended to mixed Gaussian states $\rho _{A:B}$
via the convex roof construction
\begin{equation} \label{eq:G-tangle_mixed}
\tau _{G}(\rho _{A:B})=
\inf_{\{p_{i},\rho _{i}^{(p)}\}}\sum_{i}p_{i}\tau _{G}(\rho _{i}^{(p)}),
\end{equation}
where the infimum is taken over all convex decompositions of $\rho
_{A:B}$ in terms of pure {\em Gaussian} states $\rho _{i}^{(p)}$:
$\rho _{A:B}=\sum_{i}p_{i}\rho _{i}^{(p)}$. By virtue of the convex
roof construction,  $\tau _{G}$ [Eq.~(\ref{eq:G-tangle_mixed})] is
an entanglement monotone under Gaussian LOCC (GLOCC)
\cite{WGKWC,AI05}.

Henceforth, given an arbitrary $n$-mode Gaussian state $\rho
_{A_{1}:A_{2}\ldots A_{n}}$, we refer to $\tau _{G}$
[Eq.~(\ref{eq:G-tangle_mixed})] as the {\it Gaussian tangle} and we
now prove the general monogamy inequality
\begin{equation} \label{eq:Gaussian_monogamy}
\tau _{G}(\rho _{A_{1}:A_{2}\ldots A_{n}}) \, \geq \,
\sum_{l=2}^{n}\tau _{G}(\rho _{A_{1}:A_{l}})\, .
\end{equation}

To this end, we can assume without loss of generality that the
reduced two-mode states $\rho _{A_{1}:A_{l}}=\mathrm{Tr}
_{A_{2}\ldots A_{l-1}A_{l+1}\ldots A_{n}}\rho _{A_{1}:A_{2}\dots
A_{n}}$ of subsystems $(A_{1}A_{l})$ $(l=2,\ldots ,n)$ are all
entangled. In fact, if for instance $\rho _{A_{1}:A_{2}}$ is
separable, then $\tau _{G}(\rho _{A_{1}:A_{3}\ldots A_{n}})\leq \tau
_{G}(\rho _{A_{1}:A_{2}\ldots A_{n}})$ because the partial trace
over the subsystem $A_{2}$ is a local Gaussian operation that does
not increase the Gaussian entanglement. Furthermore, by the convex
roof construction of the Gaussian tangle, it is sufficient to prove
the monogamy inequality for any {\it pure} Gaussian state $\rho
_{A_{1}:A_{2}\ldots A_{n}}^{(p)}$ (see also Refs.~\cite{CKW,OV,AI}).
Therefore, in the following we can always assume that $\rho
_{A_{1}:A_{2}\ldots A_{n}} $ is a pure Gaussian state for which the
reduced states $\rho _{A_{1}:A_{l}}$ $(l=2,\ldots ,n)$ are all
entangled.
%We also remark that the choice of subsystem $A_1$ to be
%the focus for the decomposition of bipartite entanglement in
%Eq.~(\ref{eq:Gaussian_monogamy}) is of mere convenience, as the $n$
%modes can be always rearranged so that the reference mode is labeled
%with  index $1$. Monogamy will hold with respect to any bipartition.

Some technical preliminaries are in order. A $n$-mode  (pure or
mixed) Gaussian state $\rho$ is completely characterized by the
covariance matrix (CM) $ \gamma _{jk}=2\mathrm{Tr}[\rho
(R_{j}-d_{j})(R_{k}-d_{k})]-i(J_{n})_{jk}$, and by the displacement
vector $d_{k}=\mathrm{Tr}(\rho R_{k})$. Here $ R=(\omega
_{1}^{1/2}Q_{1},\omega _{1}^{-1/2}P_{1},\ldots ,\omega
_{n}^{1/2}Q_{n},\omega _{n}^{-1/2}P_{n})^{T}$, with $Q_{k}$ and
$P_{k}$ the canonical quadrature-phase operators (position and
momentum) for mode $k$ with energy $\omega _{k}$, and
$J_{n}=\oplus_{j=1}^{n}J_{1}$ with $J_{1}={{\ 0\ 1}\choose{-1\ 0}}$,
the symplectic matrix \cite{BL}. Every unitary transformation for
general (pure or mixed) Gaussian states $\rho \mapsto U\rho
U^{\dagger }$ is associated to a symplectic transformation $\gamma
\mapsto S\gamma S^{T}$ with $S\in
\mathrm{Sp}(2n,\mathbb{R})=\{S|SJ_{n}S^{T}=J_{n}\}$. Positivity of
the density matrix $\rho$ is expressed in terms of $\gamma$ as
\begin{equation} \label{eq:positivity}
\gamma +iJ_{n}\geq 0 \,.
\end{equation}
Under partial transposition, $\rho _{A_{1}:A_{2}\ldots A_{n}}
\mapsto \rho_{A_{1}:A_{2}\ldots A_{n}}^{T_{A_{1}}}$, the CM $\gamma$
is transformed to $\widetilde{\gamma }=F_{n}\gamma F_{n}$, where
$F_{n}=\mathrm{diag}(1,-1,1,1,\ldots ,1)$.
%From
%Eq.~(\ref{eq:positivity}), if $\gamma $ is positive, so is
%$\widetilde{\gamma }$.
% One of the most
%remarkable symplectic transformations is the Williamson
%decomposition \cite{Wil}. Namely, given any positive and symmetric
%$2n\times 2n$ matrix $A$, there always exists a symplectic
%transformation $S$ that brings it to the so-called Williamson normal
%form $A =S\mathrm{diag}(\nu _{1},\nu _{1},\ldots ,\nu _{n},\nu
%_{n})S^{T}$, with $\nu _{j}$ $(> 0)$ called symplectic eigenvalues
%of $A$, $(j=1,\ldots ,n)$. Note that $\det A$ is preserved under
%symplectic transformations because $\det S=1$ \cite{Pramana}. Since
%both $\gamma $ and $\widetilde{\gamma }$ are positive and symmetric,
%they also admit the Williamson normal forms.

We start by computing the left-hand side of
\eq{eq:Gaussian_monogamy}. Since $\rho _{A_{1}:A_{2}\ldots A_{n}}$
is a $1\times (n-1)$ pure Gaussian state, it can be transformed as
\cite{Hol,BR}
\begin{equation} \label{eq:unitary}
U\rho _{A_{1}:A_{2}\ldots A_{n}}U^{T}=\rho _{A_{1}:A_{2}^{\prime }}\otimes
\rho _{A_{3}^{\prime }}\otimes \ldots \otimes \rho _{A_{n}^{\prime }}
\end{equation}
by a local unitary transformation $U=U_{A_{1}}\otimes U_{A_{2}\ldots
A_{n}}$ without changing the amount of entanglement across the
bipartition $A_{1}:A_{2}\ldots A_{n}$. In the right-hand side of
Eq.~(\ref{eq:unitary}), $\rho _{A_{1}:A_{2}^{\prime }}$ is a pure
two-mode Gaussian state (a two-mode squeezed state) and $\rho
_{A_{l}^{\prime }}$ $(l=3,\ldots ,n)$ are vacuum states. Thus, $\tau
_{G}(\rho _{A_{1}:A_{2}\ldots A_{n}})$ is equal to $\tau _{G}(\rho
_{A_{1}:A_{2}^{\prime }})= \mathcal{N}^{2}(\rho_{A_{1}:A_{2}^{\prime
}})$. In turn, $\mathcal{N}^{2}(\rho_{A_{1}:A_{2}^{\prime}}) =
(\tilde\nu _{-}^{-1}-1)^{2}/4$ \cite{VW,ASI04}, where $\tilde\nu
_{-}$ denotes the smallest symplectic eigenvalue of
$\widetilde{\gamma }_{A_{1}:A_{2}^{\prime }}=F_{2}\gamma
_{A_{1}:A_{2}^{\prime }}F_{2}$, with $\gamma _{A_{1}:A_{2}^{\prime
}}$ being the CM of $\rho _{A_{1}:A_{2}^{\prime }}$. It is easy to
compute $\tilde\nu _{-}$; $\tilde\nu _{-}=\sqrt{\det \alpha
}-\sqrt{\det \alpha -1} $, where $\alpha=~{{\gamma_{1,1}\
\gamma_{1,2}}\choose{\gamma_{2,1}\ \gamma_{2,2}}}$  is the CM of the
single-mode reduced Gaussian state $\rho _{A_{1}}=\mathrm{Tr}
_{A_{2}\ldots A_{n}}\rho _{A_{1}:A_{2}\ldots A_{n}}$.
%The CM $\gamma$ of a $n$-mode pure Gaussian state has the form
%\begin{equation} \label{eq:Euler}
%\gamma =T^{T}Z_{n}T \, ,
%\end{equation}
%where $Z_{n}=\mathrm{diag}(z_{1},z_{1}^{-1},\ldots
%,z_{n},z_{n}^{-1})$ with $z_{j}\geq 1$ $(j=1,\ldots ,n)$, and $T\in
%\mathrm{Sp}(2n,\mathbb{R})\cap O(2n)=K(n)$. Here $O(2n)$ denotes the
%orthogonal group whose elements are $2n\times 2n$ real orthogonal
%matrices \cite{SEW}. Therefore,
%$K(n)$ is isomorphic to $U(n)$ \cite{SMD}, the
%unitary group whose elements are $n\times n$ unitary matrices. The
%elements of $T$ are thus written in terms of the elements $u_{jk}$
%of a unitary matrix $U \in U(n)$ as \cite{Hir}:
%\begin{equation} \label{eq:T}
%\begin{split}
%T_{2j-1,2k-1}=T_{2j,2k}&=\mathrm{Re}\ u_{jk}\,; \\
%T_{2j-1,2k}=-T_{2j,2k-1}&=\mathrm{Im}\ u_{jk}\,.
%\end{split}
%\end{equation}
%From Eqs.~(\ref{eq:Euler}),\,(\ref{eq:T}) we can compute $\det
%\alpha =\sum_{j,k=1}^{n}\big(
%z_{j}z_{k}+z_{j}^{-1}z_{k}^{-1}\big) \big[ (\mathrm{Re}%
%u_{j1}^{*}u_{k1})^{2}+(\mathrm{Im}u_{j1}^{*}u_{k1})^{2}\big].$
%Substituting $u_{j1}^{*}u_{k1}=\delta
%_{jk}-\sum_{l=2}^{n}u_{jl}^{*}u_{kl}$, we obtain $\det \alpha
%=\sum_{l=2}^{n}\Delta _{l}/4+1$, with
%\begin{eqnarray} \label{eq:delta}
%\Delta _{l} &=&-2\sum_{j,k=1}^{n}\left(
%z_{j}z_{k}+z_{j}^{-1}z_{k}^{-1}\right)   \nonumber \\
%&\times& \left[ (\mathrm{Re}u_{jl}^{*}u_{kl})(\mathrm{Re}u_{j1}^{*}u_{k1})+(%
%\mathrm{Im}u_{jl}^{*}u_{kl})(\mathrm{Im}u_{j1}^{*}u_{k1})\right].
%\end{eqnarray}
 The CM $\gamma$ of a $n$-mode {\em pure}
Gaussian state is characterized by the condition \cite{WGKWC} $- J_n
\gamma J_n \gamma = \id_{2n}\,,$ which implies
\begin{equation}
\det \alpha + \sum_{l=2}^{n} \det \delta_{l} = 1\,, \label{speci}
\end{equation}
where $\delta_l$ is the  matrix encoding intermodal correlations
between mode $1$ and mode $l$ in the reduced state
$\rho_{A_{1}:A_{l}}$ $(l=2,\ldots ,n)$, described by a CM
\begin{equation} \label{eq:reduced_gamma}
\gamma _{A_{1}:A_{l}}=\left(
\begin{array}{cccc}
\gamma _{1,1} & \gamma _{1,2} & \gamma _{1,2l-1} & \gamma _{1,2l} \\
\gamma _{2,1} & \gamma _{2,2} & \gamma _{2,2l-1} & \gamma _{2,2l} \\
\gamma _{2l-1,1} & \gamma _{2l-1,2} & \gamma _{2l-1,2l-1} & \gamma
_{2l-1,2l}
\\
\gamma _{2l,1} & \gamma _{2l,2} & \gamma _{2l,2l-1} & \gamma
_{2l,2l}
\end{array}
\right) =\left(
\begin{array}{cc}
\alpha  & \delta _{l} \\
\delta _{l}^{T} & \beta _{l}
\end{array}
\right).
\end{equation}
As $\rho _{A_{1}:A_{l}}$ is entangled, $\det \delta _{l}$ is
negative \cite{Sim}. It is useful to introduce the auxiliary
quantities $\Delta _{l}=-4\det \delta _{l}>0$. The Gaussian tangle
for $\rho_{A_{1}:A_{2}\ldots A_{n}}$ is then written as
\begin{equation} \label{eq:monogamy1} \tau _{G}(\rho
_{A_{1}:A_{2}\ldots A_{n}})=\frac{1}{4}(\tilde\nu
_{-}^{-1}-1)^{2}=f\left( \sum_{l=2}^{n}\Delta _{l}\right),
\end{equation}
\begin{equation} \label{eq:f_g} {\rm where}\,\,
f(t)=(g^{-1}(t)-1/2)^{2},\,\,\, {\rm with}\,\,
g(t)=\sqrt{t+4}-\sqrt{t}\,.
\end{equation}
We observe that  $f(t)/t$ is an increasing function for $t>0$
and $f(0)=0$ so  $f$ is a star-shaped function: $f(ct)\leq cf(t)$
for $c\in [0,1]$ and $t\geq 0$ \cite{Comment2}. Therefore, we have
$f(t)\leq \frac{t}{t+s}f(t+s)$ and $f(s)\leq \frac{s}{t+s}f(t+s)$
for $t,s\geq 0$ to obtain $f(t)+f(s)\leq f(t+s)$. That is, $f$ is
superadditive \cite{MO}. Hence \cite{Comment3},
\begin{equation} \label{eq:monogamy2}
f\left( \sum_{l=2}^{n}\Delta _{l}\right) \geq \sum_{l=2}^{n}f(\Delta _{l}).
\end{equation}
Each term in the right-hand side is well defined since $\Delta _{l}>0$.

We are now left to compute the right-hand side of
\eq{eq:Gaussian_monogamy}, i.e.~the bipartite entanglement in the
reduced (mixed) two-mode states $\rho_{A_{1}:A_{l}}$ $(l=2,\ldots
,n)$. We will show that the corresponding Gaussian tangle is bounded
from above by $f(\Delta _{l})$, which will therefore prove the
monogamy inequality via \eq{eq:monogamy2}. To this aim, we recall
that any bipartite and multipartite entanglement in a Gaussian state
is fully specified in terms of its CM, as the displacement vector of
first moments can be always set to zero by local unitary operations,
which preserve entanglement by definition. It is thus convenient to
express the Gaussian tangle directly in terms of the CMs. Following
Refs. \cite{WGKWC,AI05,AI,ASI06}, the definition
[Eq.~(\ref{eq:G-tangle_mixed})] for the Gaussian tangle of a mixed
Gaussian state with CM $\gamma _{A_{1}:A_{l}}$ can be rewritten as
\begin{equation} \label{eq:G-tangle_simple}
\tau _{G}(\gamma _{A_{1}:A_{l}})=\inf_{\gamma _{A_{1}:A_{l}}^{(p)}}\left\{
\tau _{G}(\gamma _{A_{1}:A_{l}}^{(p)})|\gamma _{A_{1}:A_{l}}^{(p)}\leq
\gamma _{A_{1}:A_{l}} \right\} \, ,
\end{equation}
where the infimum is taken over all CMs $\gamma
_{A_{1}:A_{l}}^{(p)}$ of pure Gaussian states such that $\gamma
_{A_{1}:A_{l}}\geq \gamma _{A_{1}:A_{l}}^{(p)}$. The quantities
$\Delta _{l}$ and $\tau _{G}(\gamma _{A_{1}:A_{l}})$ for any $l$, as
well as every single-mode reduced determinant, are
$\mathrm{Sp}(2,\mathbb{R})^{\oplus n}$-invariants. For each two-mode
partition described by \eq{eq:reduced_gamma}, we can exploit such
local-unitary freedom to put the CM $\gamma _{A_{1}:A_{l}}$ in
standard form \cite{notesform} with $\alpha = {\rm diag}\{a,\,a\}$,
$\beta_l = {\rm diag}\{b,\,b\}$, and $\delta_l = {\rm
diag}\{c_{+},\,c_{-}\}$,
%\begin{equation} \label{eq:standard}
%\gamma _{A_{1}:A_{l}}=\left(
%\begin{array}{cccc}
%a & 0 & c_{+} & 0 \\
%0 & a & 0 & c_{-} \\
%c_{+} & 0 & b & 0 \\
%0 & c_{-} & 0 & b
%\end{array}
%\right) ,
%\end{equation}
where $c_{+}\geq \left| c_{-}\right| $ \cite{Sim,Duan}. The
condition [Eq.~(\ref{eq:positivity})] for $\gamma _{A_{1}:A_{l}}$ is
thus equivalent to the following inequalities
\begin{eqnarray}&
a \geq 1,\,b \geq 1,\,ab-c_{\pm }^{2} \geq 1 \; ;  \label{eq:positivity1} \\
&\!\!\!\!\! \det \gamma
_{A_{1}:A_{l}}+1=(ab-c_{+}^{2})(ab-c_{-}^{2})+1 \geq
a^{2}+b^{2}+2c_{+}c_{-}. \quad \label{eq:positivity2}
\end{eqnarray}
Furthermore, since the state $\rho _{A_{1}:A_{l}}$ is entangled, we
have \cite{Sim}
\begin{equation} \label{eq:inseparability}
(ab-c_{+}^{2})(ab-c_{-}^{2})+1 \, < \, a^{2}+b^{2}-2c_{+}c_{-} \, .
\end{equation}
From Eqs.~(\ref{eq:positivity2}) and
(\ref{eq:inseparability}), it follows that $c_{-}<0 $. In
Eq.~(\ref{eq:G-tangle_simple}), $\tau _{G}(\gamma
_{A_{1}:A_{l}}^{(p)})=f(4\det \alpha ^{(p)}-4)$, which is an
increasing function of $\det \alpha ^{(p)}$, where $\alpha ^{(p)}$
is the first $2\times 2$ principal submatrix of $\gamma
_{A_{1}:A_{l}}^{(p)}$. The infimum of the right-hand side of
Eq.~(\ref{eq:G-tangle_simple}) is achieved by the pure-state CM
$\gamma _{A_{1}:A_{l}}^{(p)}$ (with $\gamma _{A_{1}:A_{l}}^{(p)}\leq
\gamma _{A_{1}:A_{l}}$ and $\gamma_{A_{1}:A_{l}}^{(p)}+iJ_{2}\geq
0$) that minimizes $\det \alpha ^{(p)}$. The minimum value of $\det
\alpha ^{(p)}$ is given by $\min_{0\leq \theta <2\pi }m(\theta)$
\cite{AI05}, where: $m(\theta )=1+{h_{1}^{2}(\theta )}/{h_{2}(\theta
)}\,, $ with $h_{1}(\theta )=\xi _{-}+\sqrt{\eta }\cos \theta $, and
$h_{2}(\theta ) = 2(ab-c_{-}^{2})(a^{2}+b^{2}+2c_{+}c_{-}) -({\zeta
}/{\sqrt{\eta }})\cos \theta +(a^{2}-b^{2})\sqrt{1-{\xi
_{+}^{2}}/{\eta ^{2}}}\sin \theta$. Here
\begin{eqnarray}
\label{eq:xi} \xi _{\pm }&=&c_{+}(ab-c_{-}^{2})\pm c_{-}\,, \\
\label{eq:eta} \eta &=&[a-b(ab-c_{-}^{2})][b-a(ab-c_{-}^{2})]\,, \\
\zeta&=&2abc_{-}^{3}+(a^{2}+b^{2})c_{+}c_{-}^{2} \nonumber
\\&+&[a^{2}+b^{2}-2a^{2}b^{2}]c_{-}-ab(a^{2}+b^{2}-2)c_{+}\,.
\label{eq:zeta}
\end{eqnarray}
Moreover, $m(\pi )\geq \min_{0\leq \theta <2\pi }m(\theta )$
and therefore
\begin{equation} \label{eq:monogamy3}
\tau _{G}(\gamma _{A_{1}:A_{l}})\leq f(4m(\pi )-4)=f\left( 4\
\zeta_{1}^{2}/{\zeta _{2}}\right) \, ,
\end{equation}
where $\zeta _{1}=h_{1}(\pi )$ and $\zeta _{2}=h_{2}(\pi )$.
Finally, one can prove that (see the Appendix)
\begin{equation}\label{eq:appendix}
\Delta _{l}=-4\det \delta_{l}=-4c_{+}c_{-}\geq 4\ {\zeta
_{1}^{2}}/{\zeta _{2}}\,,
\end{equation}
which, being $f(t)$ [Eq.~(\ref{eq:f_g})] an increasing function of
$t$, entails that $f(\Delta _{l})\geq f( 4\ \zeta_{1}^{2}/{\zeta
_{2}})$. Combining this with \eq{eq:monogamy3} leads to the crucial
$\mathrm{Sp}(2,\mathbb{R})^{\oplus n}$-invariant condition
\begin{equation} \label{eq:bound}
\tau _{G}(\gamma _{A_{1}:A_{l}}) \le f(\Delta _{l})\,,
\end{equation}
which holds in general for all $l=2\ldots n$ and does not rely on
the specific standard form of the reduced CMs $\gamma
_{A_{1}:A_{l}}$ \cite{notesform}. Then, recalling
Eqs.~(\ref{eq:monogamy1}), (\ref{eq:monogamy2}), and
(\ref{eq:bound}), Inequality~(\ref{eq:Gaussian_monogamy}) is
established. This completes the proof of the monogamy constraint on
CV entanglement sharing for pure $n$-mode Gaussian states
distributed among $n$ parties. As already mentioned, the proof
immediately extends to arbitrary mixed Gaussian states by the
convexity of the Gaussian tangle [Eq.~(\ref{eq:G-tangle_mixed})].
\hfill $\blacksquare$

Summarizing, we have defined the Gaussian tangle $\tau_G$, an
entanglement monotone under GLOCC, and proved that it is monogamous
for all multimode Gaussian states distributed among multiple
parties. The implications of our result are manifold. The monogamy
constraints on entanglement sharing are essential for the security
of CV quantum cryptographic schemes \cite{Cry}, because they limit
the information that might be extracted from the secret key by a
malicious eavesdropper. Monogamy is useful as well in investigating
the range of correlations in Gaussian matrix-product states of
harmonic rings \cite{gmps}, and in understanding the entanglement
frustration occurring in ground states of many-body harmonic lattice
systems \cite{frusta}, which, following our findings, may be now
extended to arbitrary states beyond symmetry constraints.

On the other hand, investigating the consequences of the monogamy
property on the structure of entanglement sharing in generic
Gaussian states along the lines of Refs. \cite{AI,ASI06}, reveals
that there exist states that maximize both the pairwise entanglement
in any reduced two-mode partition, and the residual distributed
(multipartite) entanglement obtained as a difference between the
left-hand and the right-hand side in
Eq.~(\ref{eq:Gaussian_monogamy}). The simultaneous monogamy and {\it
promiscuity} of CV entanglement (unparalleled in qubit systems),
which can be unlimited in four-mode Gaussian states \cite{unlim},
allows for novel, robust protocols for the processing and
transmission of quantum and classical information \cite{ASI06}. The
monogamy inequality [Eq.~(\ref{eq:Gaussian_monogamy})] bounds the
persistency of entanglement  when one or more nodes in a CV
communication network sharing generic $n$-mode Gaussian resource
states are traced out.

At a fundamental level, the proof of the monogamy property for all
Gaussian states paves the way to a proper quantification of genuine
multipartite entanglement in CV systems in terms of the residual
distributed entanglement. In this respect, the intriguing question
arises  whether a {\em stronger} monogamy constraint exists on the
distribution of entanglement in many-body systems, which imposes a
physical trade-off on the sharing of both bipartite and genuine
multipartite quantum correlations. It would be important to
understand whether the inequality [Eq.~(\ref{eq:Gaussian_monogamy})]
holds as well for discrete-variable qu$d$its ($2 < d < \infty$),
interpolating between qubits and CV systems. If this were the case,
the (convex-roof extended) squared negativity, which coincides with
the tangle for arbitrary states of qubits and with the Gaussian
tangle for Gaussian states of CV systems, would qualify as a
universal {\it bona fide}, dimension-independent quantifier of
entanglement sharing in all multipartite quantum systems. In such
context, a deeper investigation into the analogy between Gaussian
states with finite squeezing and effective finite-dimensional
systems, focused on the point of view of entanglement sharing, may
be worthy.

{\em Acknowledgements.---} We thank Hiroshi Imai and Alessio
Serafini for support. GA and FI acknowledge financial support from
MIUR, INFN, and CNR-INFM.

{\em Appendix.---} In this Appendix we prove the inequality in the
right-hand side of Eq.~(\ref{eq:appendix}). From Eqs.~(\ref{eq:xi})
and (\ref{eq:eta}), we have $\xi _{\pm }^{2}-\eta
=-(ab-c_{-}^{2})[(ab-c_{+}^{2})(ab-c_{-}^{2})+1-a^{2}-b^{2}\mp
2c_{+}c_{-}] $. Using
Eqs.~(\ref{eq:positivity1}--\ref{eq:inseparability}), we find $\xi
_{-}^{2}>\eta $ so that $\zeta _{1}=h_{1}(\pi )=\xi _{-}-\sqrt{\eta
}>0$ ($\xi _{-}>0$) and $\xi _{+}^{2}\leq \eta $ so that $\xi
_{+}\leq \sqrt{\eta }$. Here, Eq.~(\ref{eq:positivity1}) implies
$\xi _{+}\geq c_{+}(ab-c_{-}^{2}-1)\geq 0$. However, $\xi _{+}=0$
implies $\det \gamma _{A_{1}:A_{l}}=(ab-c_{+}^{2})(ab-c_{-}^{2})=1$,
which means that $\gamma _{A_{1}:A_{l}}$ is the CM of a pure
Gaussian state. This can be ruled out from the beginnign and
therefore we have $\xi _{+}>0$. Substituting $\sqrt{\eta }\geq \xi
_{+}$ into $\zeta _{1}=\xi _{-}-\sqrt{\eta }$, we obtain
\begin{equation} \label{eq:zeta1}
0<\zeta _{1}\leq \xi _{-}-\xi _{+}=-2c_{-}\leq 2\sqrt{-c_{+}c_{-}}=\zeta
_{1}^{\prime }.
\end{equation}
Next, from Eqs.~(\ref{eq:positivity1}) and (\ref{eq:zeta}), we observe
$\zeta +(a^{2}+b^{2})(c_{+}+c_{-})(ab-c_{-}^{2}-1)
=-(a-b)^{2}[c_{+}-c_{-}(ab-c_{-}^{2})]\leq 0 $ to obtain
$\zeta \leq -(a^{2}+b^{2})(c_{+}+c_{-})(ab-c_{-}^{2}-1)\leq 0 $.
The last inequality is again due to Eq.~(\ref{eq:positivity1}).
Hence, we obtain
\begin{eqnarray} \label{eq:zeta2}
\zeta _{2} &=&h_{2}(\pi )=2(ab-c_{-}^{2})(a^{2}+b^{2}+2c_{+}c_{-})+\zeta/\sqrt{\eta} \nonumber \\
&\geq &2(a^{2}+b^{2}+2c_{+}c_{-})+\zeta/\xi _{+}=\zeta _{2}^{\prime
} \, .
\end{eqnarray}
Here, we have used Eq.~(\ref{eq:positivity1}), the inequality
$\sqrt{\eta }\geq \xi _{+}>0$, and $a^{2}+b^{2}+2c_{+}c_{-}\geq
2(ab-c_{+}^{2})>0$. Now, we observe that: $\xi _{+}(\zeta
_{2}^{\prime }-4)=2(a^{2}+b^{2}+2c_{+}c_{-})\xi _{+}+\zeta -4\xi
_{+}=-4c_{-}+3(a^{2}+b^{2})c_{-}-2a^{2}b^{2}c_{-}+2abc_{-}^{3}-2abc_{+}
+(a^{2}+b^{2})c_{+}(ab-c_{-}^{2})+8c_{-}^{2}c_{+}+4abc_{-}c_{+}^{2}-4c_{-}^{3}c_{+}^{2}
\geq -4c_{-}[a^{2}+b^{2}+2c_{+}c_{-}-(ab-c_{+}^{2})(ab-c_{-}^{2})]
+3(a^{2}+b^{2})c_{-}-2a^{2}b^{2}c_{-}+2abc_{-}^{3}-2abc_{+}
+(a^{2}+b^{2})c_{+}(ab-c_{-}^{2})+8c_{-}^{2}c_{+}+4abc_{-}c_{+}^{2}-4c_{-}^{3}c_{+}^{2}
=-(a^{2}+b^{2})c_{-}+2a^{2}b^{2}c_{-}-2abc_{-}^{3}-2abc_{+}
+(a^{2}+b^{2})c_{+}(ab-c_{-}^{2}) \geq
-2abc_{-}+2a^{2}b^{2}c_{-}-2abc_{-}^{3}-2abc_{+}+2abc_{+}(ab-c_{-}^{2})
=2ab(c_{+}+c_{-})(ab-c_{-}^{2}-1)\geq 0$, where we have used
Eq.~(\ref{eq:positivity2}). Noting that $\xi _{+}>0$, we obtain
$\zeta _{2}^{\prime }\geq 4$. Finally, Eqs.~(\ref{eq:zeta1}) and
(\ref{eq:zeta2}), with $\zeta _{2}^{\prime }\geq 4$, yield: ${\zeta
_{1}^{2}}/{\zeta _{2}}\leq {\zeta _{1}^{\prime 2}}/{\zeta
_{2}^{\prime }}=-{4c_{+}c_{-}}/{\zeta _{2}^{\prime }}\leq
-c_{+}c_{-}\,.$ \hfill $\blacksquare$

%%%%%%%%%%%%%%%%%%%%%%%%%%%%%%%%%%%%%%%%%%%%%%%%%%%%%%%%%%%%%%%%%%%%%%%%%%%%%%%
%%
%%%%%%%%%%%%%%%%%%%%%%%%%%%%%%%%%%%%%%%%%%%%%%%%%%%%%%%%%%%%%%%%%%%%%%%%%%%%%%%

%%%%%%%%%%%%%%%%%%%%%%%%%%%%%%%%%%%%%%%%%%%%%%%%%%%%%%%%%%%%%%%%%%%%%%%%%%%%%%%
%%
%%%%%%%%%%%%%%%%%%%%%%%%%%%%%%%%%%%%%%%%%%%%%%%%%%%%%%%%%%%%%%%%%%%%%%%%%%%%%%%


\begin{thebibliography}{99}
\bibitem{notelogneg}
Monogamy proofs using the squared logarithmic negativiy hold true
for the squared negativity as well, as the latter is a convex
function of the former. The converse is false.
\bibitem{Comment2}
$f(t)$ is convex for $t\geq 0$, which also implies that $f$ is star-shaped.
\bibitem{Comment3}
If we defined the Gaussian tangle for pure Gaussian states as the
squared logarithmic negativity (contangle) like in \cite{AI,ASI06},
we would have, instead of $f(t)$ in Eq.~(\ref{eq:f_g}), the quantity
$\ln^{2}[g(t)/2]$ which lacks the star-shape property. It can be
confirmed numerically that the function $\ln^{2}[g(t)/2]$ ($t\geq
0$) is not superadditive. However this does not imply the failure of
the $n$-mode monogamy inequality for the contangle \cite{AI}, which
might be proven with different techniques than those employed here.
\bibitem{notesform} The reduced two-mode CMs $\gamma _{A_{1}:A_{l}}$
cannot be all brought simultaneously in standard form \cite{WGKWC}.
However, our argument runs as follows. We apply
$\mathrm{Sp}(2,\mathbb{R})\oplus \mathrm{Sp}(2,\mathbb{R})$
operations on subsystems $A_1$ and $A_2$ to bring $\gamma
_{A_{1}:A_{2}}$ in standard form, evaluate an upper bound on the
Gaussian tangle in this representation, and derive an inequality
between local-unitary invariants, \eq{eq:bound}, that is therefore
not relying on the specific standard form in which explicit
calculations are performed. We then repeat such computation for the
remaining matrices
 $\gamma _{A_{1}:A_{l}}$ with
$l=3\ldots n$. At each step, only a single two-mode CM is in
standard form while the other ones will be retransformed in a form
with (in general) non-diagonal intermodal blocks $\delta_{k \neq
l}$. However, the determinant of these blocks (and so $\Delta_k$)
and the corresponding two-mode entanglement in the CMs $\gamma
_{A_{1}:A_{k}}$ are preserved, so the invariant condition
\eq{eq:bound} holds simultaneously for all $l=2\ldots n$.
\end{thebibliography}
\end{document}